\documentclass[final,5p,times,twocolumn]{elsarticle}

\usepackage{epsfig}
\usepackage{lineno}
\journal{Nuclear Physics B}

\begin{document}

\begin{frontmatter}

	\author{Kousuke Negishi\corref{cor1}\fnref{TUS}}
	\ead{6217619@ed.tus.ac.jp}
	\cortext[cor1]{Corresponding author.}

\title{X-ray response evaluation in subpixel level for X-ray SOI pixel detectors}

\author[TUS]{Takayoshi~Kohmura}
\author[TUS]{Kouichi~Hagino}
\author[TUS]{Taku~Kogiso}
\author[TUS]{Kenji~Oono}
\author[TUS]{Keigo~Yarita}
\author[TUS]{Akinori~Sasaki}
\author[TUS]{Koki~Tamasawa}
\author[Kyoto]{Takeshi~Go~Tsuru}
\author[Kyoto]{Takaaki~Tanaka}
\author[Kyoto]{Hideaki~Matsumura}
\author[Kyoto]{Katsuhiro~Tachibana}
\author[Kyoto]{Hideki~Hayashi}
\author[Kyoto]{Sodai~Harada}
\author[Miyazaki]{Koji~Mori}
\author[Miyazaki]{Ayaki~Takeda}
\author[Miyazaki]{Yusuke~Nishioka}
\author[Miyazaki]{Nobuaki~Takebayashi}
\author[Miyazaki]{Shoma~Yokoyama}
\author[Miyazaki]{Kohei~Fukuda}
\author[IPNS]{Yasuo~Arai}
\author[IPNS]{Toshinobu~Miyoshi}
\author[IMSS]{Shunji~Kishimoto}
\author[AAT]{Ikuo~Kurachi}

\address[TUS]{Department of Physics, School of Science and Technology, Tokyo University of Science, 2641 Yamazaki, Noda-shi, Chiba 278-8510, Japan}

\address[Kyoto]{Department of Physics, Faculty of Science, Kyoto University, Kitashirakawa, Oiwake-cho, Sakyo-ku, Kyoto-shi, Kyoto 606-8502, Japan}

\address[Miyazaki]{Department of Applied Physics, Faculty of Engineering, University of Miyazaki, 1-1 Gakuenkibanadainishi, Miyazaki-shi, Miyazaki 889-2155, Japan}

\address[IPNS]{Institute of Particle and Nuclear Studies (IPNS), High Energy Accelerator Research Organization (KEK), 1-1 Oho, Tsukuba-shi, Ibaraki 305-0801, Japan}

\address[IMSS]{Institute of Materials Structure Science (IMSS), High Energy Accelerator Research Organization (KEK), 1-1 Oho, Tsukuba-shi, Ibaraki 305-0801, Japan}

\address[AAT]{Department of Advansed Accelarator Technologies (AAT), High Energy Accelerator Research Organization (KEK), 1-1 Oho, Tsukuba-shi, Ibaraki 305-0801, Japan}

\begin{abstract}
We have been developing event-driven SOI Pixel Detectors, named ``XRPIX'' (X-Ray soiPIXel) based on the silicon-on-insulator (SOI) pixel technology, for the future X-ray astronomical satellite with wide band coverage from 0.5\,keV to 40\,keV. XRPIX has event trigger output function at each pixel to acquire a good time resolution of a few $\mu \rm s$ and has Correlated Double Sampling function to reduce electric noises. The good time resolution enables the XRPIX to reduce Non X-ray Background in the high energy band above 10\,keV drastically by using anti-coincidence technique with active shield counters surrounding XRPIX. In order to increase the soft X-ray sensitivity, it is necessary to make the dead layer on the X-ray incident surface as thin as possible. Since XRPIX1b, which is a device at the initial stage of development, is a front-illuminated (FI) type of XRPIX, low energy X-ray photons are absorbed in the 8\,$\rm \mu$m thick circuit layer, lowering the sensitivity in the soft X-ray band. Therefore, we developed a back-illuminated (BI) device XRPIX2b, and confirmed high detection efficiency down to 2.6\,keV, below which the efficiency is affected by the readout noise. In order to further improve the detection efficiency in the soft X-ray band, we developed a back-illuminated device XRPIX3b with lower readout noise. In this work, we irradiated 2--5\,keV X-ray beam collimated to 4\,$\rm \mu m \phi$ to the sensor layer side of the XRPIX3b at 6\,$\rm \mu m$ pitch. In this paper, we reported the uniformity of the relative detection efficiency, gain and energy resolution in the subpixel level for the first time. We also confirmed that the variation in the relative detection efficiency at the subpixel level reported by Matsumura et al.\cite{1} has improved.
\end{abstract}


\end{frontmatter}

 \nolinenumbers

\section{Introduction}
\label{}
Charge-coupled Devices (CCDs) are standard imaging spectroscopic detectors for X-ray astronomical satellites. Since the readout speed of the X-ray CCD is about several seconds, it is not suitable for the observation of the celestial bodies varying in a short time of several milliseconds. Therefore, we have developed XRPIX (X-Ray soiPIXel) which is an event-driven pixel detector using SOI (Silicon On Insulator) technology as a detector for future wide band X-ray astronomical satellites[1--10].
By implementing the trigger output function in each pixel, XRPIX can selectively read out only the signal of the X-ray event and can obtain high time resolution of several microseconds.
With this high time resolution, anti-coincidence processing can be performed with other detectors, it is expected to be able to reduce non X-ray signals.

\begin{figure}[tbp]
  \begin{center}
    \includegraphics[width=0.8\hsize]{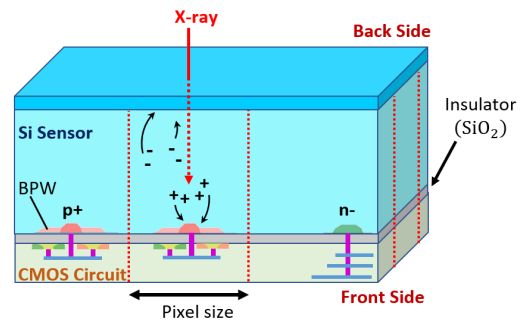}
    \caption{Schematic diagram of XRPIX when X-ray is irradiated from the back side (sensor layer side).}
    \label{fig:xrpix}
  \end{center}
\end{figure}

Figure\,\ref{fig:xrpix} shows a schematic diagram of back-illuminated (BI) type of XRPIX. 
XRPIX is a CMOS sensor having a sensor layer with a thickness of several hundreds of micrometers and a circuit layer with a thickness of about 8\,$\rm \mu m$.
A thin P layer called BPW (Buried P-Well) is formed in the sensor layer right above the CMOS circuit.
The BPW functions to suppress the back gate effect by stabilizing the potential\cite{12}.
The specifications of XRPIX are summarized in Table\,\ref{tab:hikaku}.

\begin{table*}[tbp] 
  \begin{center}
    \caption{Comparison between XRPIX1b, XRPIX2b and XRPIX3b\cite{1,10,6,15}.}
	\label{tab:hikaku}
    \begin{tabular}{l|c|c|c} \hline \hline

\multicolumn{1}{l|}{Specification} & \multicolumn{1}{c|}{XRPIX1b} & \multicolumn{1}{c|}{XRPIX2b} & \multicolumn{1}{c}{XRPIX3b}	\\ \hline \hline
depletion layer ($\rm \mu m$)	&	500 &	500 	&	300	\\
Pixel size ($\rm \mu m \times \mu m$)	&	$30.6\times$30.6	&	$30.0\times$30.0 &	$30.0\times$30.0	\\
Format (pixels)		&	$32\times32$	&	$152\times152$  &		$32\times32$	\\
Chip size ($\rm mm \times mm$)	&	$2.4\times$2.4	&	$6.0\times$6.0		&	$2.9\times$2.9	\\
Illumination type		&	FI	&	BI  &	BI	\\
Resistivity ($\rm k\Omega\cdot\rm cm$)		&	1 	&	4  &	2	\\ \hline

    \end{tabular}
  \end{center}
\end{table*} 
Since the circuit layer is composed of an 8\,$\rm \mu $m thick silicon, low energy X-rays below 1\,keV irradiated from the front side were absorbed by the circuit layer, lowering the detection efficiency. Therefore, in order to achieve high sensitivity even in the low energy band, we adopted back-illuminated type of XRPIX. In Itou et al.\,\cite{15}, we reported on the X-ray response of XRPIX2b, the first back-illuminated type of XRPIX. XRPIX2b was irradiated with X-ray of 2.6--12\,keV, and it was confirmed that the sensor layer of 500\,$\rm \mu m$ thick was fully depleted, and the quantum efficiency and the thickness of the dead layer on the back side was estimated. 
In Matsumura et al.\cite{1}, the difference of center channel (hereafter peak shift) between the single and the double pixel event due to electric field distortion was found in the XRPIX1b which was the front-illuminated (FI) type XRPIX, and it was confirmed that the peak shift was improved in the entire surface irradiation of XRPIX2b in which the pixel circuit was changed.

In this work, we have evaluated X-ray response characteristics at the subpixel level for the first time using back-illuminated device XRPIX3b which has low readout noise. In this paper, we report the evaluation of the relative detection efficiency, gain and energy resolution of subpixel level in backside irradiation of XRPIX3b.
\section{Experiment with 4\,$\rm \mu m \phi$ X-ray beams at KEK-PF}
	\subsection{Basic characteristics of XRPIX3b}
\begin{figure}[tbp]
  \begin{center}
    \includegraphics[width=0.8\hsize]{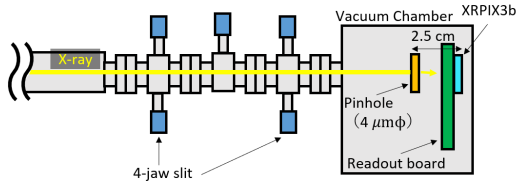}
    \caption{Schematic view from the top of experiment setup of BL-11B. XRPIX3b was irradiated with X-ray beams of 2.1\,keV and 5.0\,keV collimated to 4\,$\rm \mu m \phi$. In this experiment, we moved the stage at 6\,$\rm \mu m$ pitch and irradiated the beam at a maximum of 3$\times$3 pixels (255 spots in total).}
    \label{fig:setup}
  \end{center}
\end{figure}

XRPIX3b, being used in this research, is a device designed to improve gain and to reduce readout noise while maintaining charge collection efficiency.
XRPIX3b is the first device in which a CSA (Charge-Sensitive Amplifier) circuit is incorporated in each pixel, and a pixel circuit is laid out along BPW.
Therefore, half of the X-ray imaging area is composed of the SF (Source Follower) circuit similar to the XRPIX1b, and the other half is introducing the CSA circuit.
As the gain increased due to the introduction of the CSA circuit, we succeeded in spectroscopy of Mn-K$\alpha$ and Mn-K$\beta$ of ${}^{55}\rm Fe$ for the first time in the XRPIX series in Takeda et al.\cite{6}.
In XRPIX3b, the readout noise was 35\,$\rm e^-$\,(rms), and the energy resolution was 320\,eV in FWHM at 5.9\,keV\cite{6}.
In addition, the device used in this work was made by the FZ (Floating Zone) method, and the resistivity is 2\,$\rm k\Omega\cdot\rm cm$.
	\subsection{Experimental setup}
In order to evaluate the X-ray response at the subpixel level of the device, it is indispensable to use an X-ray beam which is monochromatic and parallel light.
We used synchrotron radiation to satisfy these conditions.
In April 2017, we conducted experiments with beamline BL-11B in synchrotron radiation facility (KEK-PF) of High Energy Accelerator Research Organization.
In BL-11B, we can use monochromatic light in the soft X-ray region by using a focusing mirror and a high vacuum double-crystal monochromator\cite{13}.
In this experiment, we irradiated X-rays of 2.1\,keV and 5.0\,keV using Si double-crystal monochromator.
\begin{figure}[tbp]
  \begin{center}
    \includegraphics[width=0.8\hsize]{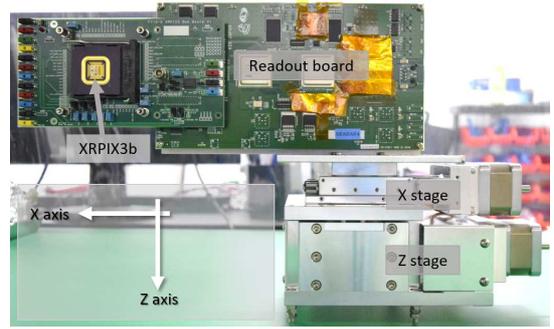}
    \caption{Readout board and XZ stage. The device is mounted on a black square table on the left side of the readout board. We plug the Ethernet cable into the right side of the readout board and read out the data. The XZ stage can move in the X direction and the Z direction. In this way, we shifted the irradiation position of the X-ray at a pitch of 6 $\rm \mu m$.}
    \label{fig:stage}
  \end{center}
\end{figure}

Figure\,\ref{fig:setup} shows a schematic diagram of experiment setup.
We attached two 4-jaw slits downstream of the beamline BL-11B, and  connected a vacuum chamber to the end.
Inside the chamber, we installed a pinhole and XRPIX3b to align with the beam axis.
Here, the pinhole and the device were spaced about 2.5 $\rm cm$ apart.
Figure\,\ref{fig:stage} shows the readout board and XZ stage used in this experiment.
Inside the vacuum chamber, the readout board is attached to a stage that can slide in the X direction and the Z direction as shown in Figure\,\ref{fig:stage}, and the device attached to the readout board can also be moved in the X direction and the Z direction at a few $\rm \mu m$ scale interval.
In this experiment, we measured the subpixel level X-ray response characteristics of XRPIX3b by moving the XZ stage at 6 $\rm \mu m$ pitch, shifting the irradiation point of the beam.
We shaped the beam using two 4-jaw slits and introduced the beam into the pinhole.
The pinhole was 4 $\mu \rm m \phi$ in diameter and 90 $\mu \rm m$ in thickness, and the material of the membrane was gold.

During the measurement, the degree of vacuum in the vacuum chamber was about 8.6$\times 10^{-7}$\,$\rm hPa$ and the temperature of the device was $-70 {}^\circ\mathrm{C}$.
The XRPIX series has two readout modes, a frame mode and an event-driven mode. The former reads out all pixels after a periodic exposure interval, and the latter reads out only a fired pixel or the inclusive peripheral pixels specified just after the triggering.
In this experiment, data was acquired in frame mode in order to read out not only specific data but all pixels.
We applied a back bias voltage of 350\,V to the sensor layer.
The exposure time per frame was 300\,$\rm \mu s$ for 5.0\,keV and 900\,$\rm \mu s$ for 2.1\,keV.
Since the intensity of the beam was greater at 5.0\,keV X-ray than at 2.1\,keV, the exposure time per frame is different in order to avoid pile-up.
We irradiated the beam to CSA region of XRPIX3b.
The X-rays were penetrated through the back side (sensor layer side) of the device. For 5.0\,keV X-ray, the number of measurement spots over $3\times3$ pixel area was 225. For 2.1\,keV X-ray, the number of measurement spots over approximate 1 pixel area was 55.
The number of frames per one spot was 150000 frames for both 5.0\,keV and 2.1\,keV.

\section{Data analysis and result}
	\subsection{Event selection} 
	\label{sec:ev}

When analyzing the data, we first preset event threshold and split threshold.
When the output value of one pixel is larger than the output values of the neighboring 8 pixels and the preset event threshold, we extract that event as an X-ray event.
Next, we make a multi-pixel event determination.
A multi-pixel event is an event in which the generated charge is detected across multiple pixels.
When an electron cloud is generated at a position distant from the circuit layer, it widely spreads until it reaches the readout nodes on the surface of the sensor layer.
In this case, the charge spreads over multiple pixels and is more likely to be detected as a multi-pixel event.
Because the charges are collected at multiple readout nodes in a multi-pixel event, it is necessary to sum up the charges in order to measure the energy of the incident X-rays.
Multiple pixel events are judged depending on the number of adjacent pixels, whose output exceed the split threshold.
We define single pixel event as the event in that electric charges are collected within one pixel, and double pixel event as the event in that the charges are collected within two pixels.
Double pixel event sum up the charges of two pixels.
The total pixel events is the sum of all those multi-pixel events.

In this work, we determined event threshold based on data acquired without irradiating X-ray beam in KEK-PF.
When acquiring data without irradiating X-rays, with an event threshold of 13\,ADU at least, no signal exceeded the event threshold in 150000 frames.
The readout noise in this experiment was 33\,$\rm e^-$, which almost agreed with Takeda et al.\cite{6}.
Since the standard deviation of zero peak was 2.3 ADU, the split threshold was set to 7\,ADU which is 3$\sigma$ so that the contamination of pedestal becomes about 1\,\% or less.
From the above reasons, we decided the event threshold as 13\,ADU and the split threshold as 7\,ADU corresponding to 0.7\,keV and 0.4\,keV, respectively, and performed data analysis.
	\subsection{Determination of pixel boundary}
\begin{figure}[tbp] 
  \begin{center}
    \includegraphics[width=0.8\hsize]{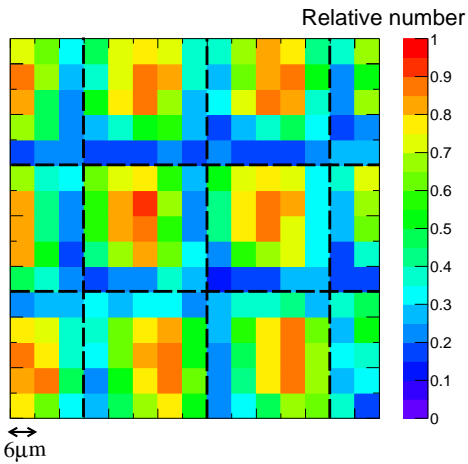} 
    \caption{Two-dimensional map of the relative number of single pixel events against the number of total pixel events at 5.0\,keV. The black dotted lines are a place presumed to be a pixel boundary. Each area is spaced 6\,$\rm \mu m$ at a time.}
    \label{fig:Grade}
  \end{center}
\end{figure}

Figure\,\ref{fig:Grade} shows the fraction of single pixel event at 5.0\,keV.
The single pixel event is an event in which the charge generated by the X-ray is collected within one pixel. Therefore, when the X-rays hit in the centers of the pixels, the ratio of the single pixel events to the total pixel events becomes maximum, and when the X-rays hit in the four pixel boundaries, this ratio becomes minimum.
In Figure\,\ref{fig:Grade}, the areas dominated by the single pixel event are located periodically, and are surrounded by the areas with a low occurrence probability of single pixel event.
These areas are repeated at a distance of about 1 pixel pitch, that is, at intervals of 30\,$\mu \rm m$.
Therefore, we considered the area with the lowest fraction of the single pixel event as pixel boundaries.
The black dotted lines are the boundary positions.

	\subsection{Uniformity of the relative detection efficiency}
\begin{figure*}[tbp] 
    \begin{tabular}{cc} 

	\begin{minipage}{0.5\hsize}
		\begin{center}
			\includegraphics[height=60mm]{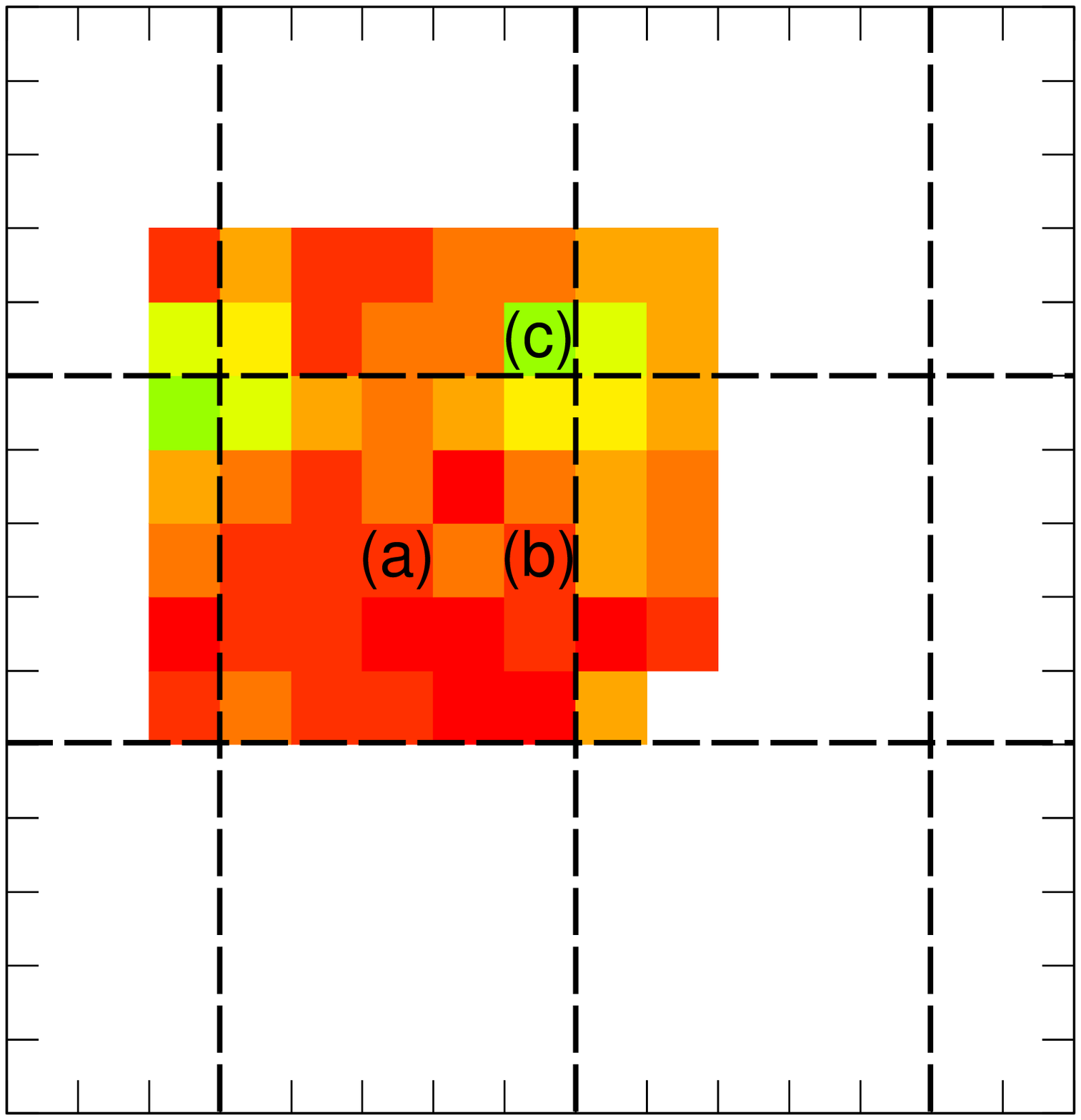} 
		\end{center}
	\end{minipage} 

	\begin{minipage}{0.5\hsize}
		\begin{center}
			\includegraphics[height=61mm]{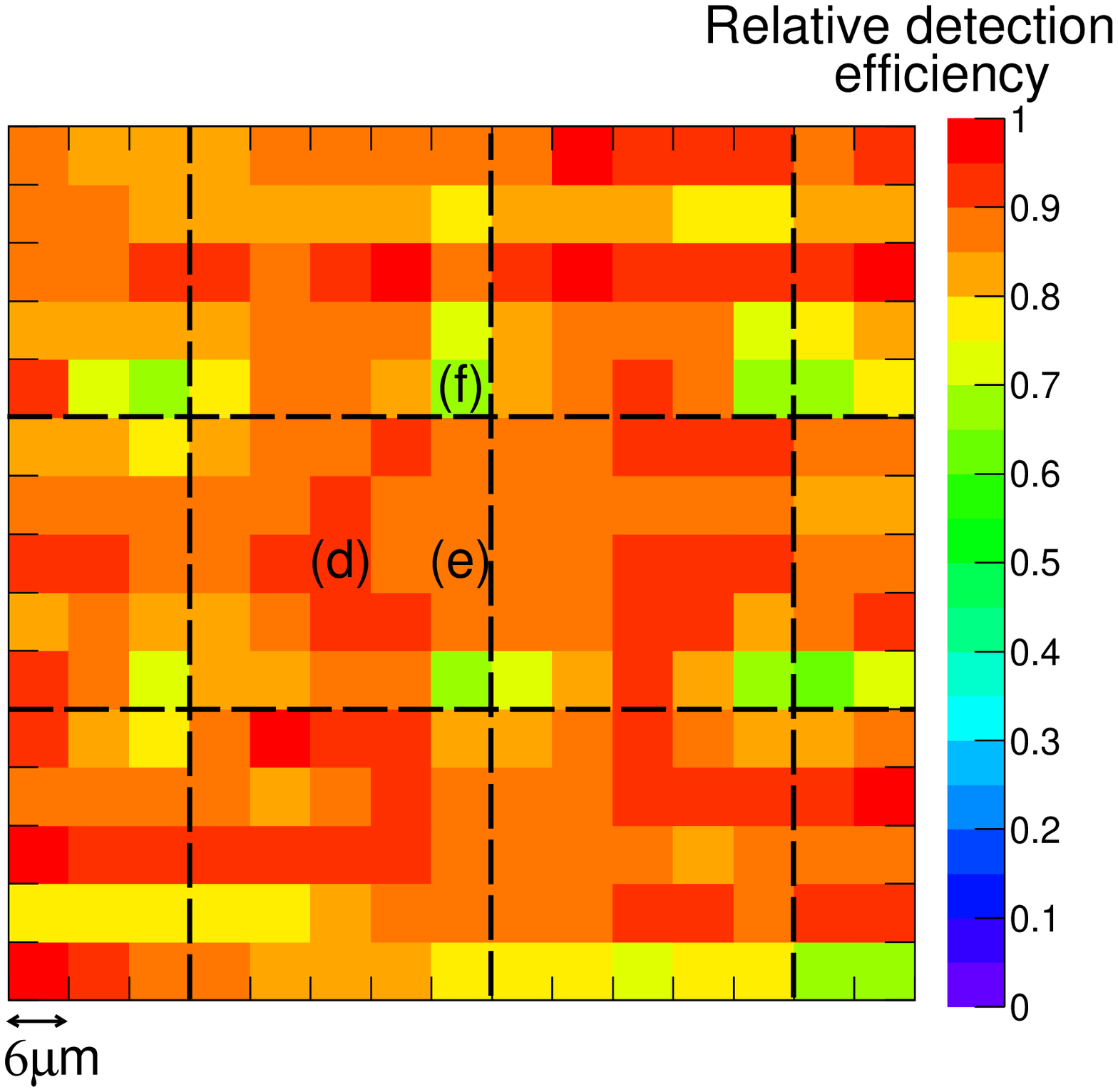} 
		\end{center}
	\end{minipage}

	\end{tabular}
	\caption{Two-dimensional map of relative detection efficiency. The left panel is a two-dimensional map of the photon count obtained when each position of the pixel is irradiated with an X-ray beam of 2.1\,keV. The right panel is 5.0\,keV. We standardized the obtained photon counts with the largest counts and display it as a two dimensional map. The black dotted line indicates the pixel boundary, and the interval is 30 $\rm \mu m$.}
	\label{fig:map}
\end{figure*}

Figure\,\ref{fig:map} shows the two-dimensional map of the X-ray count rate obtained from the beam experiments.
We normalized the photon counts obtained at the different X-ray hit positions, and plotted them as a two-dimensional map in Figure\,\ref{fig:map}. The left and the right plots show the photon counts for 2.1\,keV X-ray and 5.0\,keV, respectively.
In addition, the black dotted lines indicate the position of the pixel boundary, which are identical to the lines in Figure\,\ref{fig:Grade}.
The interval between the dotted lines are 30\,$\mu \rm m$.
Area (a) and area (d) in Figure\,\ref{fig:map} correspond to the pixel center, area (b) and area (e) correspond to the two pixel boundary, and area (c) and area (f) correspond to the four pixel boundary, respectively.

In the case of irradiating 2.1\,keV X-rays, the relative detection efficiency of the area (b) to the area (a) is $99.0\pm4.4\%$ and the relative detection efficiency of the area (c) to the area (a) is $74.0\pm3.2\%$.
Similarly for 5.0\,keV X-ray irradiation, the relative detection efficiency of the area (e) to the area (d) was $95.7\pm2.2\%$ and the relative detection efficiency of the area (f) to the area (d) was $76.3\pm1.9\%$.

	\subsection{Energy spectrum in subpixel level}

We also extracted the energy spectrum of each area.
Figure\,\ref{fig:5keV_spec_2D} shows a two-dimensional map of 5.0\,keV spectrum folded in one pixel.
The horizontal axis of the spectrum is channel and the vertical axis is the number of counts.
The shape of the spectrum is sharper at the center of the pixel, and single pixel event is dominant.
The proportion of the double pixel event is large at the two pixel boundary, and the spectrum is collapsed at the four pixel boundary.

The peak shift is seen at two pixel boundary and four pixel boundary.
In Figure\,\ref{fig:5keV_spec_2D}, the peak positions of the single pixel event and the double pixel event are nearly equal in the left and right areas of the two pixel boundary, but the peak shift can be conspicuously seen in the areas around the upper and the lower two pixel boundaries.
This data suggests that there is difference in charge collection between two cases: dividing the charges vertically, and dividing the charge horizontally. The cause for this is under investigation.
The peak of the double pixel event was shifted by about 10\% with respect to that of the single pixel event.

	\subsection{Uniformity of gain and energy resolution}

\begin{figure*}[tbp] 
  \begin{center}
    \includegraphics[width=0.7\hsize]{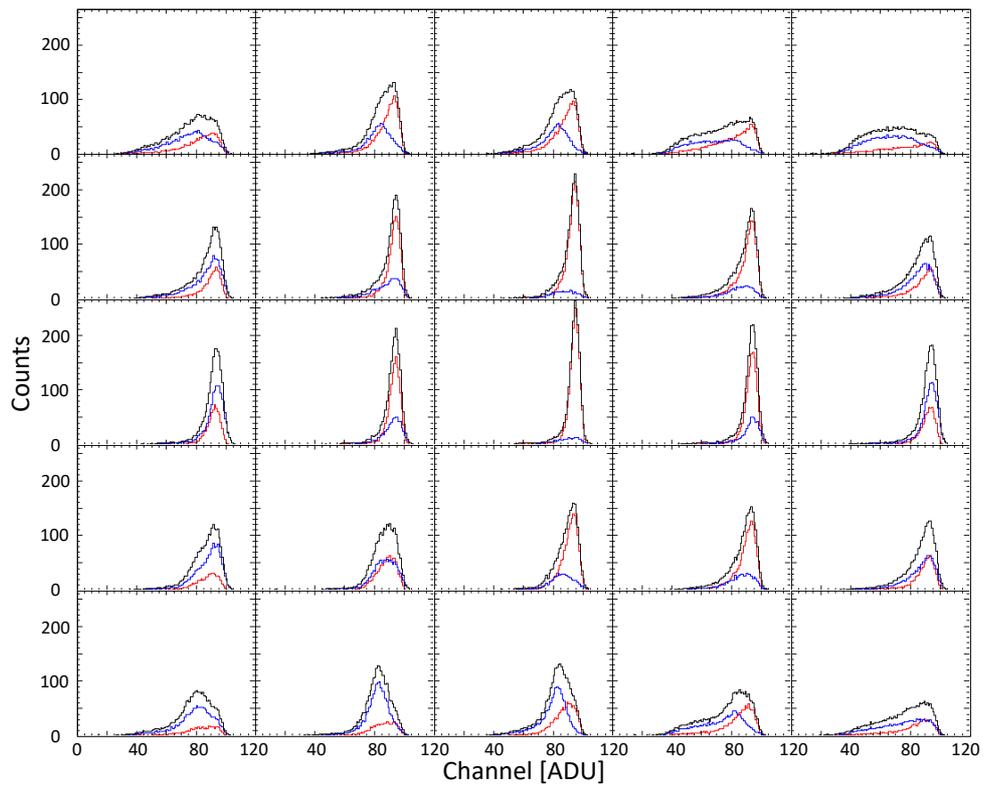}
    \caption{Two-dimensional map of the energy spectrum folded in one pixel at 5.0\,keV. The histograms of red, blue, and black represent the single, double, and total pixel event, respectively. We folded the spectrum of the whole area of Figure\,\ref{fig:map} into one pixel.}
    \label{fig:5keV_spec_2D}
  \end{center}
\end{figure*}

\begin{figure}[tbp] 
  \begin{center}
    \includegraphics[width=0.8\hsize]{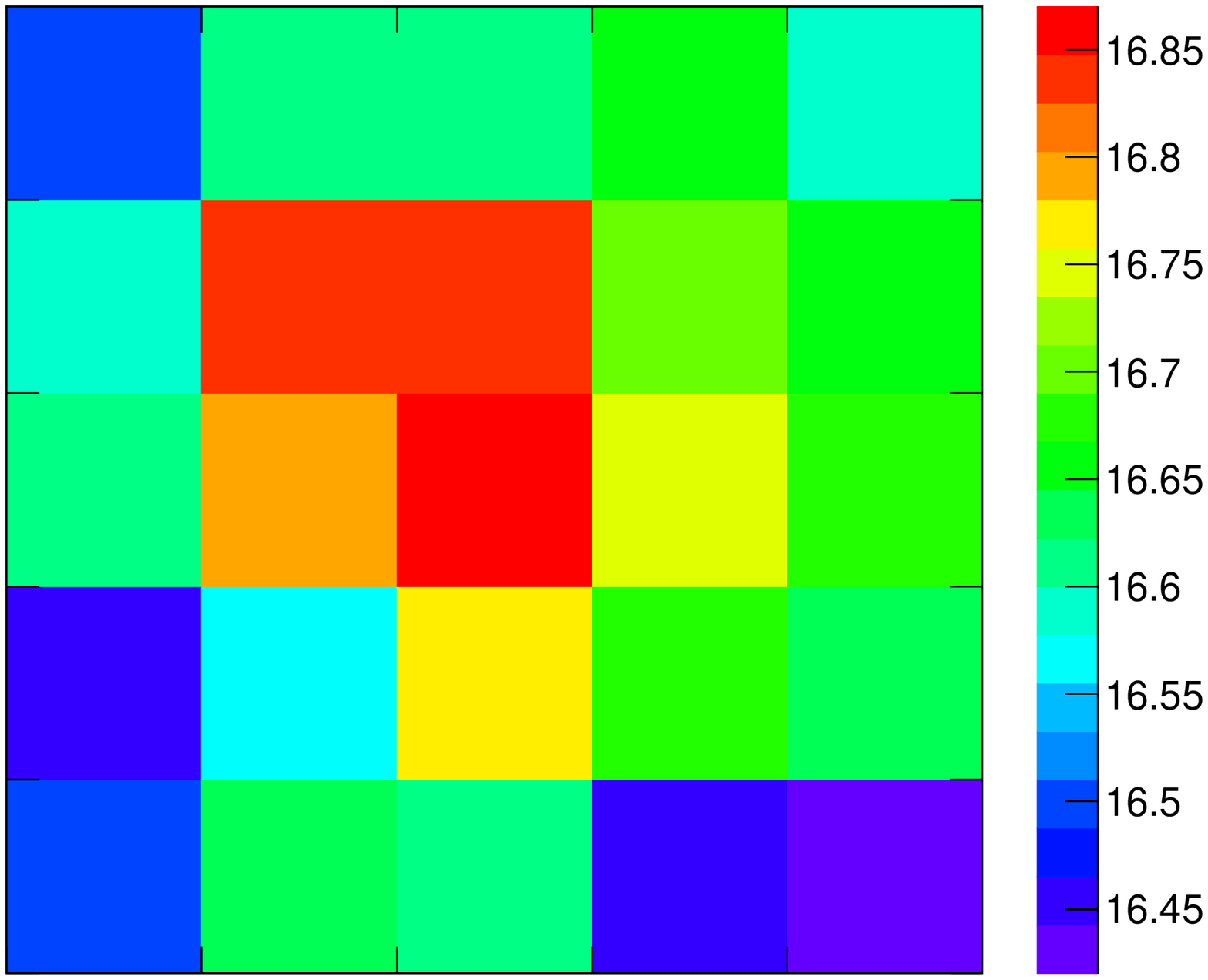}
    \caption{Two-dimensional map of gain[$\rm \mu V/e^-$] for 5.0\,keV X-ray irradiation. The value in each pixel in this map was derived from the single pixel spectrum in corresponding panel in Figure\,\ref{fig:5keV_spec_2D}. The color bar on the right side shows the magnitude of the gain.}
    \label{fig:gain_2D}
  \end{center}
\end{figure}

\begin{figure}[tbp] 
  \begin{center}
    \includegraphics[width=0.85\hsize]{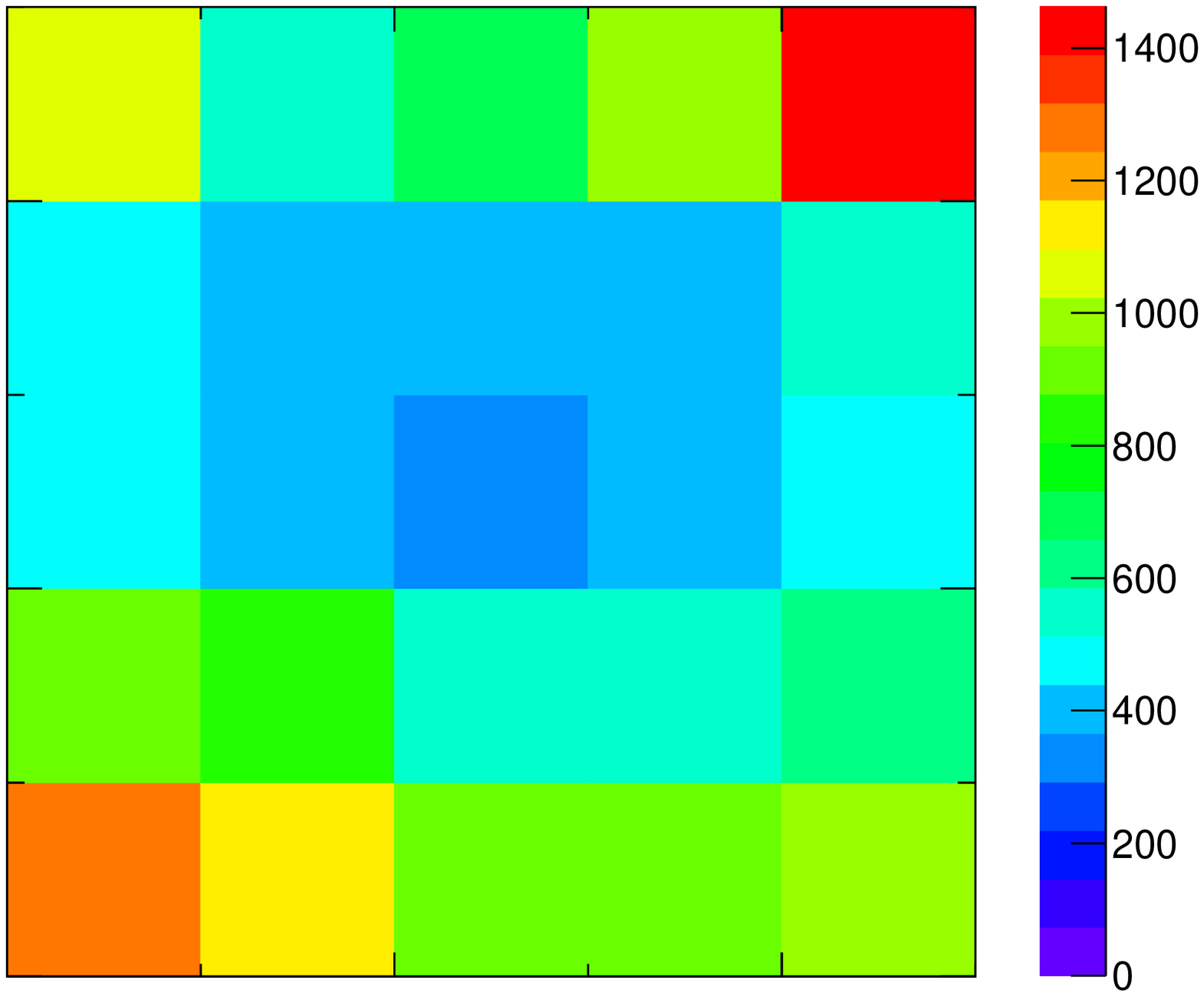}
    \caption{Two-dimensional map of energy resolution[eV] for 5.0\,keV X-ray irradiation. The value in each pixel in this map was derived from the single pixel spectrum in corresponding panel in Figure\,\ref{fig:5keV_spec_2D}. The color bar on the right side shows the magnitude of the energy resolution.}
    \label{fig:FWHM_2D}
  \end{center}
\end{figure}

We examined the uniformity of gain and energy resolution for 5.0\,keV X-ray irradiation. The value in each pixel in Figure\,\ref{fig:gain_2D} and \ref{fig:FWHM_2D} was derived from the single pixel spectrum in corresponding panel in Figure\,\ref{fig:5keV_spec_2D}.

Figure\,\ref{fig:gain_2D} shows a two-dimensional map of gain.
The color bar on the right side of the two-dimensional map of Figure\,\ref{fig:gain_2D} represents the magnitude of the gain, and the unit is $\rm \mu V/e^-$.
The gain was derived from the center value of the main gaussian by using double gaussian fitting for the single pixel event.
The sub gaussian represents the tail structure of the spectrum.
It can be seen that the value of the gain is the largest at the center of one pixel and the lowest at the four pixel boundary from Figure\,\ref{fig:gain_2D}. The highest gain is 16.866$\pm$0.006\,$\rm \mu V/e^-$.
The area with the lowest gain is the lower right area, but the deviation from the pixel center was within about 2.6\%.


Figure\,\ref{fig:FWHM_2D} shows a two-dimensional map of energy resolution in FWHM at 5.0\,keV.
As with the gain, we derived it from the single pixel event.
The color bar on the right side represents the magnitude of the energy resolution, and the unit is the electron volt.
It can be seen that the energy resolution is better at the pixel center.
This is also evident from Figure\,\ref{fig:5keV_spec_2D}.
The area with the highest energy resolution was the area at the center of Figure\,\ref{fig:FWHM_2D}, which was 362$\pm$2\,eV in FWHM at 5.0\,keV.

We summed up the spectra of all irradiated areas of 5.0\,keV and found the gain and energy resolution in the same condition as irradiating X-rays to all the areas.
As a result, gain was 16.774$\pm$0.004\,$\rm \mu V/e^-$, and energy resolution was 374$\pm$2\,eV.

\section{Discussion and Summary}

We have been developing XRPIX for wide energy band X-ray imaging spectroscopy from 0.5 to 40\,keV. In the XRPIX3b used in this work, the energy resolution has been improved to 350\,eV by increasing the output gain, and the readout noise is lower than XRPIX2b in order to realize high sensitivity\cite{15}. The detection efficiency of the back side of XRPIX2b and the thickness of the dead layer were evaluated in the Itou et al.\cite{15}. In this work, we evaluated the uniformity of the relative detection efficiency, the gain, and the energy resolution at the subpixel level for the first time by irradiating the XRPIX3b with the X-ray to the back side. We irradiated this device with 2.1\,keV and 5.0\,keV X-rays collimated to 4\,$\rm \mu m$ at KEK-PF.

\begin{figure}[tbp] 
  \begin{center}
    \includegraphics[width=0.8\hsize]{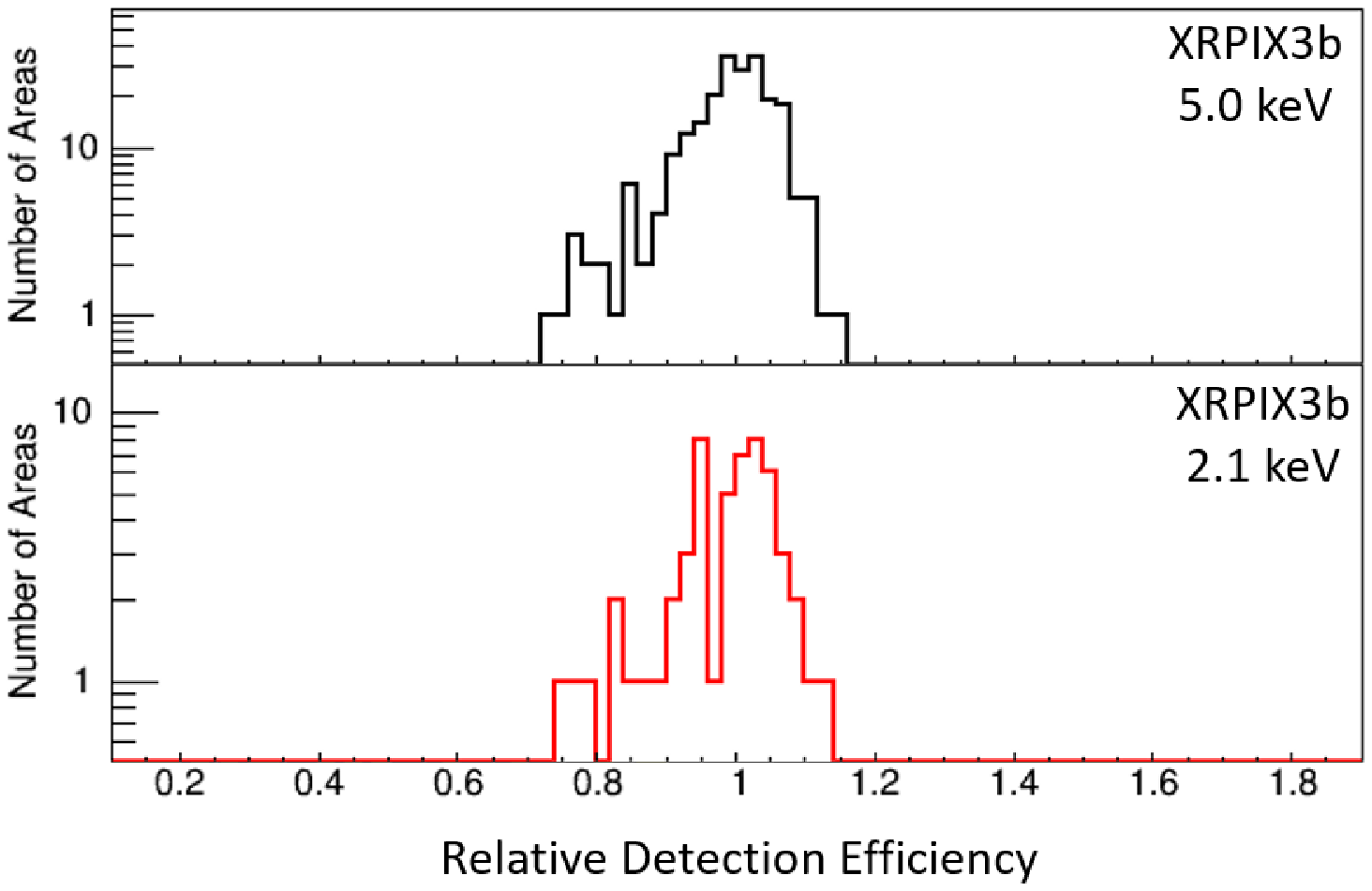}
    \caption{Variation of relative detection efficiency. The number of detections in each area of Figure\,\ref{fig:map} was divided by the median of all areas and displayed as a histogram. The upper and lower panel shows the variation of relative detection efficiency of 5.0\,keV and 2.1\,keV of XRPIX3b, respectively. The vertical axis represents the number of areas irradiated with X-rays, and the horizontal axis represents variation in the number of detections of each area with respect to the median of all areas.}
    \label{fig:5keV_QE_hist}
  \end{center}
\end{figure}
\begin{figure}[tbp]
  \begin{center}
    \includegraphics[width=0.8\hsize]{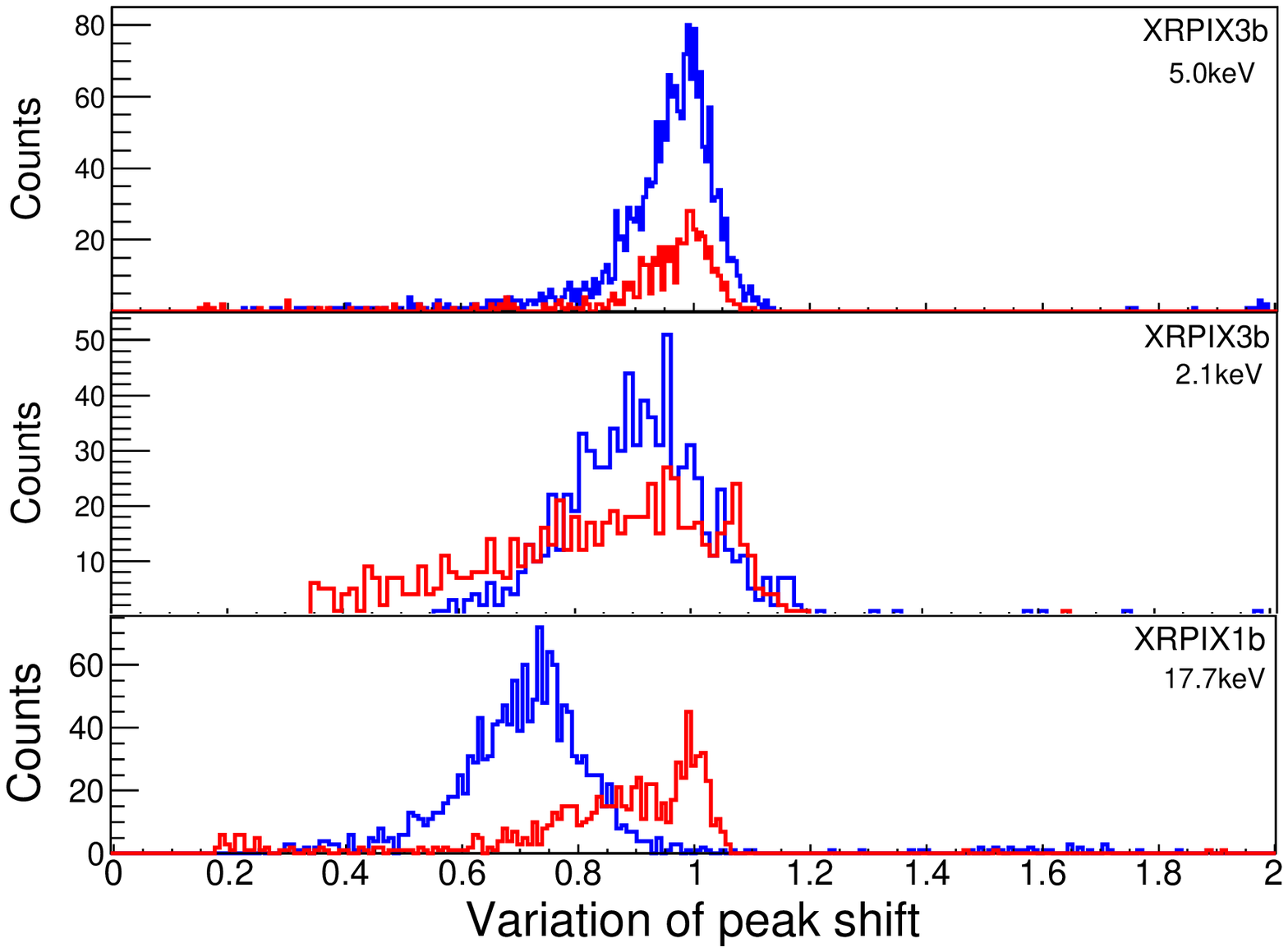} 
    \caption{Energy spectrum at two pixel boundary with peak position aligned to 1. Spectra of 5.0\,keV and 2.1\,keV of XRPIX3b, and 17.7\,keV of XRPIX1b in order from the top panel. Red and blue histogram show single and double pixel event, respectively.}
    \label{fig:peakshift}
  \end{center}
\end{figure}

Figure\,\ref{fig:5keV_QE_hist} shows the variation of the relative detection efficiency. A fitting of the histogram of XRPIX3b at 5.0\,keV with a Gaussian function provides the standard deviation of 5.0$\pm$0.3\%. The peak-to-peak variation of the relative detection efficiency was about 43\% at 5.0\,keV and 38\% at 2.1\,keV for XRPIX3b. 

In addition, Figure\,\ref{fig:peakshift} shows 4 spectra in which the center channel of single pixel event at two pixel boundary are set to 1 to compare the peak shifts between the single and the double pixel event. 
The magnitude of the peak shifts are about 1.8\% at 5.0\,keV and 9.5\% at 2.1\,keV for XRPIX3b, and about 29.0\% at 17.7\,keV for XRPIX1b.
In Matsumura et al.\cite{1}, the peak shift found in XRPIX1b was confirmed to be improved in XRPIX2b by X-ray irradiation on the entire imaging area of XRPIX2b. In this experiment, we confirmed that the peak shift is improved also in subpixel scale in XRPIX3b.
 
We also evaluated the uniformity of gain and energy resolution in one pixel.
The gain at the pixel center area was 16.866$\pm$0.006\,$\rm \mu V/e^-$, and the difference between center area and the lowest area was about 2.6\%.
The energy resolution at the pixel center area was 362$\pm$2\,eV in FWHM at 5.0\,keV, and the four pixel boundary area which had worst energy resolution was about 4 times higher value in FWHM than its center area.

Compared to XRPIX1b, the relative detection efficiency within one pixel and the center channel shift in energy spectrum was improved, then we found the charge collection efficiency was improved in XRPIX3b.
Since we can expect further improvements with the latest device that we are developing, we would like to investigate the device in the future.

\section*{Acknowledgments}
We acknowledge the valuable advice and great work by the personnel of LAPIS Semiconductor Co., Ltd. This study was supported by the Japan Society for the Promotion of Science (JSPS) KAKENHI Grant-in-Aid for Scientific Research on Innovative Areas 25109002 (Y.A.), 25109003 (S.K.), 25109004 (T.G.T., T.T. and T.K.), Grant-in-Aid for Scientific Research (B) 25287042 (T.K.), Grant-in-Aid for Young Scientists (B) 15K17648 (A.T.), Grant-in-Aid for Challenging Exploratory Research 26610047 (T.G.T.) and Grant-in-Aid for JSPS Fellows 15J01842 (H.M.). This study was also supported by the VLSI Design and Education Center (VDEC), the University of Tokyo in collaboration with Cadence Design Systems, Inc., and Mentor Graphics, Inc.

\end{document}